\documentclass[twocolumn,showpacs]{revtex4}
\usepackage{eurosym}
\usepackage{graphicx}
\usepackage{dcolumn}
\usepackage{bm}
\usepackage{amsmath}
\usepackage{amsfonts}
\usepackage{amssymb}

\providecommand{\U}[1]{\protect\rule{.1in}{.1in}}

\begin{document}

\title{Measurements in the L\'{e}vy quantum walk}
\author{A. Romanelli}
\altaffiliation{\textit{E-mail address:} alejo@fing.edu.uy}
\affiliation{Instituto de F\'{\i}sica, Facultad de Ingenier\'{\i}a\\
Universidad de la Rep\'ublica\\
casilla de correo 30, c\'odigo postal 11000, Montevideo, Uruguay}
\date{\today }

\begin{abstract}
We study the quantum walk subjected to measurements with a L\'evy
waiting-time distribution. We find that the system has a sub-ballistic
behavior instead of a diffusive one. We obtain an analytical expression for
the exponent of the power law of the variance as a function of the
characteristic parameter of the L\'evy distribution.
\end{abstract}

\maketitle

\section{Introduction}

The development of the quantum walk (QW) in the context of quantum
computation, as a generalization of the classical random walk, has attracted
the attention of researchers from different fields. The fact that it is
possible to build and preserve quantum states experimentally has led the
scientific community to think that quantum computers could be a reality in
the near future. On the other hand, from a purely physical point of view,
the study of quantum computation allows to analyze and verify the principles
of quantum theory. In this last frame the study of the QW subjected to
different sources of decoherence is a topic that has been considered by
several authors \cite{kendon}. In particular we have recently studied \cite%
{alejo0} the QW and the quantum kicked rotor in resonance subjected to noise
with a L\'evy waiting-time distribution \cite{Levy}, finding that both
systems have a sub-ballistic wave function spreading, as shown by the
power-law tail of the standard deviation ($\sigma (t)\sim t^{c}$ with $%
0.5<c<1$), instead of the known ballistical growth ($\sigma (t)\sim
t$). This sub-ballistic behavior was also observed in the dynamics
of both the quantum kicked rotor \cite{alejo3} and the QW
\cite{Ribeiro} when these systems are subjected to an excitation
that follows an aperiodic Fibonacci prescription. Other authors also
investigated the kicked rotor subjected to noises with a L\'{e}vy
distribution \cite{Shomerus} and almost-periodic Fibonacci sequence
\cite{Casati}, showing that this decoherence never fully destroys
the dynamical localization of the kicked rotor but leads to a
sub-diffusion regime for a short time before localization appears.
All these mentioned papers have in common that they work with
quantum systems that have an anomalous behavior that was established
numerically. There are no analytical results that explain in a
general way why noises, with a power-law distribution or in a
Fibonacci sequence, lead the system to a new non-diffusive behavior.
Here we present a simple model that allows an analytical treatment
to understand the sub-ballistic behavior. We hope that this may help
to understand in a generical way how the frequency of the
decoherence is the main factor in this unexpected dynamics. With
this aim we investigate the QW when measurements are performed on
the system with waiting times between them following a L\'{e}vy
power-law distribution. We show that this noise produces a change
from ballistic to sub-ballistic behavior and we obtain analytically
a relation between the exponent of the standard deviation and the
characteristic parameter of L\'{e}vy distribution.

The paper is organized as follows. In the next section we develop the QW
model with L\'{e}vy noise, in the third section analytical results are
obtained and in the last section we draw the conclusions.

\section{Quantum walk and measurement}

\label{walked}The dynamics of the QW subjected to a series of measurements
will be generated by a large sequence of two time-step unitary operators $%
U_{0}$ and $U_{1}$ as was done in a previous work \cite{alejo0}. But now $%
U_{0}$ is the `free' evolution of the QW and $U_{1}$ is the operator that
measures simultaneously the position and the chirality of the QW. The time
interval between two applications of the operator $U_{1}$ is generated by a
waiting-time distribution $\rho (T) $, where $T$ is a dimensionless integer
time step. The detailed mechanism to obtain the evolution is given in \cite%
{alejo0}. We take $\rho (T)$ in accordance with the L\'{e}vy distribution
\cite{Shlesinger,Klafter,Zaslavsky} that includes a parameter $\alpha $,
with $0<\alpha \leq 2$. When $\alpha< 2$ the second moment of $\rho$ is
infinite, when $\alpha=2$ the Fourier transform of $\rho$ is the Gaussian
distribution and the second moment is finite. Then, this distribution has no
characteristic size for the temporal jump, except in the Gaussian case. The
absence of scale makes the L\'{e}vy random walks scale-invariant fractals.
This means that any classical trajectory has many scales but none in
particular dominates the process. The most important characteristic of the L%
\'{e}vy noise is the power-law shape of the tail, accordingly in this work
we use the waiting-time distribution
\begin{equation}
\rho (t)=\frac{\alpha }{\left( 1+\alpha \right) }\left\{
\begin{array}{cc}
1, & 0\leq t<1 \\
\left( \frac{1}{t}\right) ^{\alpha +1}, & t\geq 1%
\end{array}%
\right. .  \label{Levy}
\end{equation}%
To obtain the time interval $T$ we sort a continuous variable $t$ in
agreement with eq. (\ref{Levy}) and then we take the integer part $T_{i}$ of
this variable \cite{alejo0}.

To obtain the operator $U_{0}$ we develop in some detail the free QW model.
The standard QW corresponds to a one-dimensional evolution of a quantum
system (the walker) in a direction which depends on an additional degree of
freedom, the chirality, with two possible states: `left' $|L\rangle $ or
`right' $|R\rangle $. Let us consider that the walker can move freely over a
series of interconnected sites labeled by an index $n$. In the classical
random walk, a coin flip randomly selects the direction of the motion; in
the QW the direction of the motion is selected by the chirality. At each
time step a rotation (or, more generally, a unitary transformation) of the
chirality takes place and the walker moves according to its final chirality
state. The global Hilbert space of the system is the tensor product $%
H_{s}\otimes H_{c}$ where $H_{s}$ is the Hilbert space associated to the
motion on the line and $H_{c}$ is the chirality Hilbert space.

If one is only interested in the properties of the probability distribution
it suffices to consider unitary transformations which can be expressed in
terms of a single real angular parameter $\theta $ \cite{Nayak,Tregenna,Bach}%
. Let us call $M_{-}$ ($M_{+}$) the operators that move the walker one site
to the left (right) on the line in $H_{s}$ and let $|L\rangle \langle L|$
and $|R\rangle \langle R|$ be the chirality projector operators in $H_{c}$.
Then we consider free evolution transformations of the form \cite{Nayak},
\begin{equation}
U_{0}(\theta )=\left\{ M_{-}\otimes |L\rangle \langle L|+M_{+}\otimes
|R\rangle \langle R|\right\} \circ \left\{ I\otimes K(\theta )\right\} ,
\label{Ugen}
\end{equation}%
where $K(\theta )=\sigma _{z}e^{-i\theta \sigma _{y}}$ is an unitary
operator acting on $H_{c}$, $\sigma _{y}$ and $\sigma _{z}$ being the
standard Pauli matrices, and $I$ is the identity operator in $H_{s}$. The
unitary operator $U_{0}(\theta )$ evolves the state $|\Psi (t)\rangle $ by
one time step,
\begin{equation}
|\Psi (t+1)\rangle =U_{0}(\theta )|\Psi (t)\rangle .  \label{evol1}
\end{equation}%
The wave-vector $|\Psi (t)\rangle $ is expressed as the spinor
\begin{equation}
|\Psi (t)\rangle =\sum\limits_{n=-\infty }^{\infty }{\binom{a_{n}(t)}{%
b_{n}(t)}}|n\rangle ,  \label{spinor}
\end{equation}%
where we have associated the upper (lower) component to the left (right)
chirality, the states $|n\rangle $ are eigenstates of the position operator
corresponding to the site $n$ on the line. The unitary evolution for $|\Psi
(t)\rangle $, corresponding to eq.~(\ref{evol1}) can then be written as the
map
\begin{align}
a_{n}(t+1)& =a_{n+1}(t)\,\cos \theta +b_{n+1}(t)\,\sin \theta \,,
\label{mapa} \\
b_{n}(t+1)& =a_{n-1}(t)\,\sin \theta -b_{n-1}(t)\,\cos \theta \,.  \notag
\end{align}%
To build the measurement operator $U_{1}$ we consider the case in which the
walker starts from the position eigenstate $|0\rangle $, and with an initial
qubit state $(a_{0},b_{0})=(1,i)/\sqrt{2}$. The operator $U_{1}$ must
describe the measurement of position and chirality simultaneously. The
measurement of position is direct, but among the many ways to measure the
chirality we choose to do it in such a way that the two qubit states $(1,i)/%
\sqrt{2},$ and $(1,-i)/\sqrt{2}$ are eigenstates of the measurement
operator. This means that we project the chirality on the $y$ direction
using the $\sigma _{y\text{ }}$ Pauli operator. This form of measurement
ensures that the initial conditions after each measurement are equivalent to
each other from the point of view of the probability distribution \cite%
{konno}. In this work we take $\theta =\pi /4$ as in the usual Hadamard walk
on the line. The probability distribution for the walker's position at time $%
t$ is given by
\begin{equation}
P_{n}(t)=|a_{n}(t)|^{2}+|b_{n}(t)|^{2}.  \label{prob}
\end{equation}%
The effect of performing measurements on this system at time
intervals $T_{i} $, with a L\'{e}vy distribution, combined with
unitary evolution is as follows: the standard deviation has a
ballistic growth during the free evolution, and when a measurement
collapses the wave function the standard deviation starts again from
zero. Then the standard deviation has a zig-zag path.
\begin{figure}[th]
\begin{center}
\includegraphics[scale=0.42]{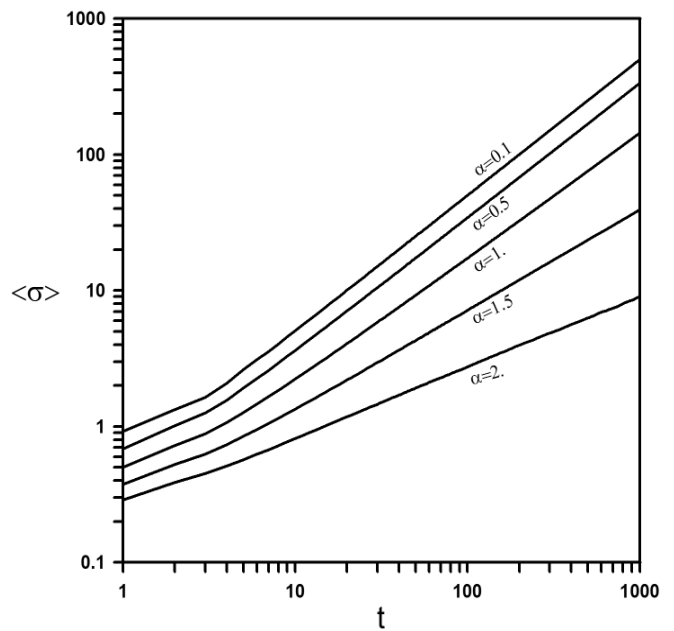}
\end{center}
\caption{The standard deviation for the QW as a function of dimensionless
time in logarithmic scales. The parameters of the curves from top to bottom
are: (1) $\protect\alpha =0.1$ and $c=1.$; (2) $\protect\alpha =0.5$ and $%
c=0.99$; (3) $\protect\alpha =1.$ and $c=0.92$; (4) $\protect\alpha =1.5$
and $c=0.74$ and (5) $\protect\alpha =2.$ and $c=0.51$. }
\label{fig1}
\end{figure}
In Fig.~\ref{fig1} we present the numerical calculation of the average
standard deviation $\left\langle \sigma \left( t\right) \right\rangle $ of
the QW with measurements, calculated through a computer simulation of the
time evolution of an ensemble of $2\times 10^{6}$ stochastic trajectories
for each value of the parameter $\alpha $. This figure shows for different
values of $\alpha $ that the behavior is diffusive ($\sim t$) for times $%
t\lesssim 10$ and follows a power law $\sim t^{c}$ for times $t\gtrsim 10$.
One would expect that the large degree of decoherence introduced by
measurements led to a diffusive behavior for all times \cite{alejo0};
instead the system changes to an unexpected sub-ballistic regime. Then $%
\alpha $ determines the degree of diffusivity of the system for large
waiting times, that is ballistic for $\alpha =0$, sub-ballistic for $%
0<\alpha <2$ and diffusive for $\alpha =2$. \label{walked} Fig.~\ref{fig2}
shows, in full line, the exponent $c$ of the power law as a function of the L%
\'{e}vy parameter $\alpha $. Again the calculation has been made with an
ensemble of $2\times 10^{6}$ stochastic trajectories for each value of $%
\alpha $. The passage from the ballistic behavior for $\alpha =0$ to the
diffusive behavior for $\alpha =2$ is clearly observed. In this figure the
result of the analytical calculation of the next section is also presented
in dashed line, the comparison of the curves proving the coherence of both
treatments.

\section{Theoretical model}

\label{sec:derivation} Different mechanisms of unitary noise may
drive a system from a quantum behavior at short times to a
classical-like one, at longer times. It is clear that the quadratic
growth in time of the variance of the QW is a direct consequence of
the coherence of the quantum evolution \cite{alejo1}. In this
section, we develop an analytical treatment to understand why the
L\'{e}vy decoherence does not fully break the ballistic behavior.

In our previous work \cite{alejo2} we investigated the QW on the
line when decoherence was introduced through simultaneous
measurements of the chirality and position. In that work it was
proved that the QW shows a diffusive behavior when measurements are
made at periodic times or with a Gaussian distribution. Now, we use
the L\'{e}vy distribution but some of the result obtained in
\cite{alejo2} can be used. For the sake of clarity we reproduce
briefly the main steps to obtain the dynamical equation of the
variance. Let us suppose that the wave-function is measured at the
time $t$, then it evolves according to the unitary map~(\ref{mapa})
during a time interval $T$, and again at this last time $t+T$, a new
measurement is performed. The probability that the wave-function
collapses in the eigenstate $|n\rangle $ due to the position
measured, after a time $T$, is
\begin{equation}
q_{n}\equiv P_{n}(T)\,.  \label{position}
\end{equation}
These spatial distributions $q_{n}$ depend on the initial qubit state and
the time interval $T$, they will play the role of transition probabilities
for the global evolution. The mechanism used to perform measurements of
position and chirality assures that these distributions will repeat
themselves around the new position, because the initial chirality ($(1,i)/%
\sqrt{2}$ or $(1,-i)/\sqrt{2}$) produces the same spatial distributions $%
q_{n}$ and their value only depends on the size of $T$. Then it is
straightforward to build the probability distribution $P_{n}$ at the
new time $t+T$ as a convolution between this distribution at the
time $t$ with the conditional probability $q_{n}(T)$; this takes the
form of the following `master equation'
\begin{equation}
P_{n}(t +T)=\sum\limits_{j=n-T}^{n+T}q_{n-j}P_{j}(t),  \label{markov1}
\end{equation}
where $q_{n-j}$ are the transition probabilities from site $j$ to site $n$,
defined in eq.~(\ref{position}) and the sum is extended between $j=n-T$ and $%
j=n+T$, because $T$ is also used as the number of applications of the
quantum map eq.~(\ref{mapa}). Remember that for each time step the walker
moves one spatial step to the right and left. Using eq.~(\ref{markov1}) we
calculate the first moment $m_{1}(t)\equiv\sum jP_{j}(t)$ and the second
moment $m_{2}(t)\equiv\sum j^{2}P_{j}(t)$ to obtain
\begin{align}
m_{1}(t+T)& =m_{1}(t)+m_{1q}(T)  \label{mom1} \\
m_{2}(t+T)& =m_{2}(t)+2m_{1}(t)m_{1q}(T)+m_{2q}(T)  \label{mom2}
\end{align}
where ${m_{1q}(T)=\sum\limits_{n=-T}^{n=T}nq_{n}}$ and ${m_{2q}(T)=
\sum\limits_{n=-T}^{n=T}n^{2}q_{n}}$ are the first and second moments of the
unitary evolution between measurements.
\begin{figure}[th]
\begin{center}
\includegraphics[scale=0.38]{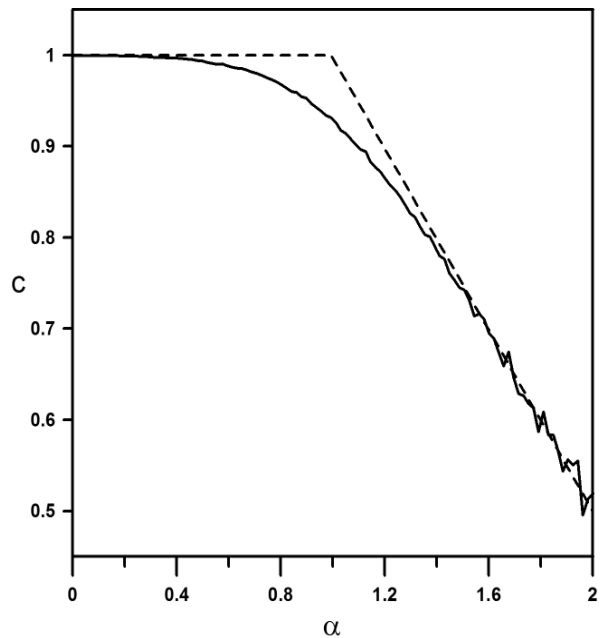}
\end{center}
\caption{the exponent $c$ of the power law of the standard deviation for the
QW as a function of the parameter $\protect\alpha$. The full line correspond
to the numerical result and the dashed line to the analytical one}
\label{fig2}
\end{figure}
Therefore the global variance $\sigma ^{2}(t)=m_{2}(t)-m_{1}^{2}(t)$
verifies the following equation for the process, obtained for first time in
\cite{alejo2}
\begin{equation}
\sigma ^{2}\left( t+T\right) =\sigma ^{2}\left( t\right) +\sigma _{q}^{2}(T),
\label{varianza}
\end{equation}%
where $\sigma _{q}^{2}(T)=m_{2q}(T)-m_{1q}^{2}(T)$ is the variance
associated to the unitary evolution between measurements. These results can
be used for random time intervals $T$ between consecutive measurements. The
variance $\sigma _{q}^{2}$ depends very weakly on the qubit's initial
conditions and it increases quadratically with time \cite{Nayak}
\begin{equation}
\sigma _{q}^{2}(T)=kT^{2}\,,  \label{varianza2}
\end{equation}%
where $T\gg 1$ and $k$ is a constant determined by the initial
conditions. We now calculate the average of eq. (\ref{varianza})
taking integer time intervals $T_{i}$ between consecutive
measurements sorted according to the L\'{e}vy distribution eq.
(\ref{Levy})
\begin{equation}
\left\langle \sigma ^{2}\left( t+T_{i}\right) \right\rangle =\left\langle
\sigma ^{2}\left( t\right) \right\rangle +k\left\langle
T_{i}^{2}\right\rangle ,  \label{varianza3}
\end{equation}%
where $\left\langle f(t)\right\rangle \equiv {\int\limits_{0}^{t}f(x){\rho
(x)}dx}$. From the previous numerical calculation, we know that the variance
grows as a power law, then $\left\langle \sigma ^{2}\left( t\right)
\right\rangle \propto t^{2c}$ and $\left\langle \sigma ^{2}\left(
t+T_{i}\right) \right\rangle \propto \left\langle( t+T_{i})
^{2c}\right\rangle\approx t^{2c}(1+2c\frac{\left\langle T_{i}\right\rangle}{t%
})$ for large $t$. Substituting these expressions in eq.~(\ref{varianza3}),
the following result for the exponent $c$ is obtained
\begin{equation}
c\approx\frac{1}{2}\left( 1+\frac{\log \frac{\left\langle
T_{i}^{2}\right\rangle }{\left\langle T_{i}\right\rangle }}{\log t}\right) .
\label{exponente}
\end{equation}%
valid for a large $t$. The first and the second moments of the waiting time
of our L\'{e}vy distribution are
\begin{equation}
\left\langle T_{i}\right\rangle =\frac{\alpha }{\alpha +1}\left\{ 1+\frac{%
t^{1-\alpha }-1}{1-\alpha }\right\} ,  \label{first}
\end{equation}%
\begin{equation}
\left\langle T_{i}^{2}\right\rangle =\frac{\alpha }{\alpha +1}\left\{ \frac{1%
}{3}+\frac{t^{2-\alpha }-1}{2-\alpha }\right\} .  \label{secon}
\end{equation}%
Therefore, in the case when $t\rightarrow \infty $ the exponent $c$ is%
\begin{equation}
c=\left\{
\begin{array}{cc}
1 \, , & \text{if \ }0\leqslant \alpha \leqslant 1 \\
\frac{1}{2}(3-\alpha ) \, , & \text{if \ }1\leqslant \alpha \leqslant 2.%
\end{array}%
\right.  \label{limite}
\end{equation}%
This result is in accordance with the numerical result obtained in the
previous section, as can be seen in Fig.~\ref{fig2}. It is important to
remark that eq.~(\ref{limite}) gives the analytical dependence of the
exponent of the power law for $\left\langle \sigma ^{2}\right\rangle$ on the
parameter $\alpha $.

\section{Conclusion}

\label{sec:conclusion} Several systems have been proposed as candidates to
implement the QW model, they include atoms trapped in optical lattices \cite%
{Dur,Roldan}, cavity quantum electrodynamics \cite{Sanders} and
nuclear magnetic resonance in solid substrates \cite{Du,Berman}. All
these proposed implementations face the obstacle of decoherence due
to environmental noise and imperfections. Thus the study of the QW
subjected to different types of noise may be important in future
technical applications. Here we showed that the QW subjected to
measurements with a L\'{e}vy waiting-time distribution does not
break completely the coherence in the dynamics, but produces a
sub-ballistic behavior in the system, as an intermediate situation
between the ballistic and the diffusive behavior. Note that as
Gaussian noise is a particular case of the L\'{e}vy noise, our study
is open to wider experimental situations. We studied this behavior
numerically and we obtained also an analytical expression for the
exponent of the power law of the variance as a function of the
characteristic parameter of the L\'evy distribution. Measurement is
one of the strongest possible decoherences on any quantum system,
but even in this case the coherence of the QW treated here is not
fully lost . This fact shows that the most important ingredient for
the sub-ballistic behavior is the temporal sequence of the
perturbation, not its size or type. This extreme and simple model
that can
be treated analytically allows us to understand why similar models \cite%
{alejo0,Ribeiro,alejo3} (that were studied numerically) show also a
power-law behavior.

\section{Acknowledgments}

\label{sec:Acknowledgments} I thank V. Micenmacher and R. Siri for their
comments and stimulating discussions. I acknowledge the support from
PEDECIBA and PDT S/C/IF/54/5.

\end{document}